\documentclass[12pt]{article}

\catcode`@=12
\begin{document}
\thispagestyle{empty}
\begin{flushright} 
UCRHEP-T339\\ 
June 2002\
\end{flushright}
\vspace{0.5in}
\begin{center}
{\LARGE	\bf Heavy Triplet Leptons and New Gauge Boson\\}
\vspace{1.5in}
{\bf Ernest Ma$^a$ and D. P. Roy$^{a,b}$\\}
\vspace{0.2in}
{$^a$ Physics Department, University of California, Riverside, 
California 92521, USA\\}
{$^b$ Tata Institute of Fundamental Research, Mumbai (Bombay) 400005, India\\}
\vspace{1.5in}
\end{center}
\begin{abstract}\
A heavy triplet of leptons $(\Sigma^+, \Sigma^0, \Sigma^-)_R$ per family is 
proposed as the possible anchor of a small seesaw neutrino mass.  A new 
U(1) gauge symmetry is then also possible, and the associated gauge boson 
$X$ may be discovered at or below the TeV scale.  We discuss the 
phenomenology of this proposal, with and without possible constraints 
from the NuTeV and atomic parity violation experiments, which appear to 
show small discrepancies from the predictions of the standard model.
\end{abstract}

\newpage
\baselineskip 24pt

\section{Introduction}

To obtain nonzero neutrino masses so as to explain the observed atmospheric 
\cite{atm} and solar \cite{sol} neutrino oscillations, the minimal standard 
model of particle interactions is often extended to include three neutral 
fermion singlets, usually referred to as right-handed singlet neutrinos.  If 
they have large Majorana masses, then the famous seesaw mechanism 
\cite{seesaw} allows 
the observed neutrinos to acquire naturally small Majorana masses. 
On the other hand, there are other equivalent ways \cite{masa,ma98} to realize 
this effective dimension-five operator \cite{wein} for neutrino mass. 
For example, if we replace each neutral fermion singlet by a 
triplet: \cite{ma98,foot}
\begin{equation}
\Sigma = (\Sigma^+, \Sigma^0, \Sigma^-) \sim (1,3,0)
\end{equation}
under $SU(3)_C \times SU(2)_L \times U(1)_Y$, the seesaw mechanism works 
just as well.

If the Majorana mass of $\Sigma$ is very large, then its effect at low 
energies is indistinguishable from that of the canonical seesaw.  On the 
other hand, if it is at or below the TeV energy scale, which is a natural 
possibility if there exists a second Higgs doublet, as shown recently 
\cite{ma01}, then there are interesting new experimental signatures for 
the origin of neutrino mass.  In Section 2, the phenomenology of this 
scenario is discussed.

It is well-known \cite{mamo} that in the case of one additional right-handed 
singlet neutrino per family of quarks and leptons, it is possible to promote 
$B-L$ (baryon number -- lepton number) from being a global U(1) symmetry to 
an U(1) gauge symmetry.  Similarly a new $U(1)_X$ gauge symmetry \cite{ma02} 
is also possible here.  The model is described in Section 3.

Since the $X$ gauge boson may be at or below the TeV scale, it may be 
responsible for some of the possible discrepancies observed in recent 
experiments.  In Section 4, we use it to explain the NuTeV result \cite{nutev} 
and explore its phenomenological implications.  In Section 5, we do the same 
but using atomic parity nonconservation \cite{bewi} as a constraint.  In 
Section 6, the Higgs sector is discussed and its difference from other 
proposals is noted.  We then conclude in Section 7.

\section{Heavy Triplet Leptons}

Instead of the usual singlet $N_R$, consider the addition of a fermion triplet 
$(\Sigma^+, \Sigma^0, \Sigma^-)_R$ per family to the particle content of the 
standard model.  The Yukawa interaction
\begin{equation}
{\cal L}_Y = f [-\phi^- \bar \nu_L \Sigma^+_R + {1 \over \sqrt 2} (\bar \phi^0 
\bar \nu_L + \phi^- \bar e_L) \Sigma^0_R + \bar \phi^0 \bar e_L \Sigma^-_R] + 
h.c.
\end{equation}
means that a seesaw neutrino mass matrix is obtained, i.e.
\begin{equation}
{\cal M}_{\nu \Sigma} = \left[ \begin{array} {c@{\quad}c} 0 & f v/\sqrt 2 \\ 
f v/\sqrt 2 & M_\Sigma \end{array} \right],
\end{equation}
together with $e-\Sigma$ mixing \cite{foot}, i.e.
\begin{equation}
{\cal M}_{e \Sigma} = \left[ \begin{array} {c@{\quad}c} m_e & f v \\ 0 & 
M_\Sigma \end{array} \right].
\end{equation}
For $v = \langle \phi^0 \rangle = 174$ GeV as in the standard model, either 
$M_\Sigma$ has to be very large or $f$ very small for this to give realistic 
neutrino masses.  However, if a symmetry exists which replaces $\Phi$ in 
Eq.~(2) with a second Higgs doublet $\eta = (\eta^+, \eta^0)$ having a very 
small vacuum expectation value, then $M_\Sigma$ may be at or even below the 
TeV scale.  Such a model has already been described \cite{ma01}, using $N_R$. 
It is straightforward to apply it here in the context of $\Sigma_R$.

The idea is very simple.  Assign lepton number $L = -1$ to $\eta$ but 
$L=0$ to $\Phi$ and $\Sigma$, then $M_\Sigma$ is an allowed Majorana mass, 
and $\Phi$ is replaced with $\eta$ in Eq.~(2).  Let $L$ be broken by 
explicit $soft$ terms in the Lagrangian, i.e. $\mu_{12}^2 \Phi^\dagger \eta 
+ h.c.$, then for $m_\eta^2 > 0$ and large,
\begin{equation}
\langle \eta^0 \rangle = u \simeq -{ \mu_{12}^2 v \over m_\eta^2}.
\end{equation}
For $\mu_{12}^2 \sim 10$ GeV$^2$, $v \sim 10^2$ GeV, $m_\eta \sim 1$ TeV, 
we get $u \sim 1$ MeV and $m_\nu \simeq f^2 u^2/2M_\Sigma \sim 1$ eV or less 
if $M_\Sigma \sim 1$ TeV.  Note that after the explicit breaking of $L$, a 
residual symmetry is still conserved, i.e. the conventional multiplicative 
lepton number, where $\nu$, $e$, and $\Sigma$ are odd, but $\Phi$ and $\eta$ 
are even.  In other words, there are no unwanted $\Delta L = 1$ interactions 
even though $\langle \eta^0 \rangle \neq 0$.

Whereas $\Sigma^0$ has no coupling to either the photon or the $Z$ boson 
(as is the case with $N_R$), $\Sigma^\pm$ interacts with both.  Hence our 
proposal is more easily tested experimentally than the canonical seesaw.  
If the mass of $\Sigma^\pm$ is below that of $\eta$, the former decays only 
via its mixing with $e^\pm$.  Thus we expect the decay modes
\begin{equation}
\Sigma_i^+ \to e_j^+ Z, ~\bar \nu_j W^+, ~~~ \Sigma_i^- \to e_j^- Z, 
~\nu_j W^-,
\end{equation}
which would map out the Yukawa coupling matrix $f_{ij}$, leading to a 
good determination of the neutrino mass matrix itself \cite{ma01}, up to 
an overall scale.  On the other hand, if $M_\Sigma > m_\eta$, then the decays
\begin{equation}
\Sigma^+_i \to e^+_j \eta^0, ~\bar \nu_j \eta^+
\end{equation}
will also determine $f_{ij}$.  The subsequent decays of $\eta^+$ and 
$\eta^0$ occur through their small mixings with $\phi^+$ and $\phi^0$, so 
they are dominated by $t \bar b$ and $t \bar t$ final states and should be 
easily identifiable.

Since $\Sigma$ and $\eta$ have distinctive signatures once they are produced, 
their discoveries are primarily controlled by the size of the signal.  We 
have estimated their pair production cross sections at the LHC and at the 
Tevatron via the standard Drell-Yan mechanism.  The spin and color averaged 
matrix element squares are given by
\begin{equation}
\overline {M^2}_{q \bar q \to \eta^+ \eta^-} = {1 \over 3} e^4 (ut-m_\eta^4) 
\left[ \left( {Q_q \over s} + {L_q L_\eta \over s-M_Z^2} \right)^2 + 
\left( {Q_q \over s} + {R_q L_\eta \over s-M_Z^2} \right)^2 \right],
\end{equation}
where
\begin{equation}
L_\eta = {{1 \over 2} - \sin^2 \theta_W \over \sin \theta_W \cos \theta_W}, 
~~~ L_q = {I_q^3 - Q_q \sin^2 \theta_W \over \sin \theta_W \cos \theta_W}, 
~~~ R_q = {-Q_q \sin^2 \theta_W \over \sin \theta_W \cos \theta_W}.
\end{equation}
The analogous matrix element squares for $\Sigma^\pm$ pair production are
\begin{equation}
\overline {M^2}_{q \bar q \to \Sigma^+ \Sigma^-} = {1 \over 3} e^4 \left[ 
(t-M_\Sigma^2)^2 \left( {Q_q \over s} + {L_q R_\Sigma \over s-M_Z^2} \right)^2 
+ (u-M_\Sigma^2)^2 \left( {Q_q \over s} + {R_q R_\Sigma \over s-M_Z^2} 
\right)^2 \right],
\end{equation}
where
\begin{equation}
R_\Sigma = {1-\sin^2 \theta_W \over \sin \theta_W \cos \theta_W}.
\end{equation}
Figure 1 shows the LHC and Tevatron production cross sections of the heavy 
scalar pair $\eta^\pm$ and Figure 2 shows those of the heavy lepton pair 
$\Sigma^\pm$ as functions of their mass.  We see from these figures that the 
final luminosity of about 300 $fb^{-1}$ at the LHC will correspond to a modest 
discovery limit of both $\eta^\pm$ and $\Sigma^\pm$ up to a mass of about 1 
TeV.

\section{New Gauge Boson}

Consider $SU(3)_C \times SU(2)_L \times U(1)_Y \times U(1)_X$ as a possible 
extension of the standard model, under which each family of quarks and 
leptons transforms as follows:
\begin{eqnarray}
&& (u,d)_L \sim (3,2,1/6;n_1), ~~~ u_R \sim (3,1,2/3;n_2), ~~~ 
d_R \sim (3,1,-1/3;n_3), \nonumber \\ 
&& (\nu,e)_L \sim (1,2,-1/2;n_4), ~~~ e_R \sim (1,1,-1;n_5), ~~~ 
\Sigma_R \sim (1,3,0;n_6).
\end{eqnarray}
It has been shown recently \cite{ma02} that $U(1)_X$ is free of all 
anomalies \cite{ava,mixed,witten} for the following assignments:
\begin{equation}
n_2 = {1 \over 4} (7 n_1 - 3 n_4), ~~ n_3 = {1 \over 4} (n_1 + 3 n_4), 
~~ n_5 = {1 \over 4} (-9 n_1 + 5 n_4), ~~ n_6 = {1 \over 4} (3 n_1 + n_4).
\end{equation}
This is a remarkable and highly nontrivial result.

As shown in Ref.~[10], there are 6 conditions to be satisfied for the gauging 
of $U(1)_X$.  Three of them do not involve $n_6$ and have 2 solutions:
\begin{eqnarray}
({\rm I}): && n_3 = 2n_1-n_2, ~~ n_4 = -3n_1, ~~ n_5 = -2n_1-n_2; \\ 
({\rm II}): && n_2 = {1 \over 4} (7n_1-3n_4), ~~ n_3 = {1 \over 4} 
(n_1+3n_4), ~~ n_5 = {1 \over 4} (-9n_1+5n_4).
\end{eqnarray}
The other 3 involve $n_6$, and they are given by
\begin{eqnarray}
{1 \over 2} (3n_1 + n_4) &=& {1 \over 3}p(p+1)(2p+1) n_6, \\ 
6n_1 - 3n_2 - 3n_3 + 2n_4 - n_5 &=& (2p+1) n_6, \\ 
6n_1^3 - 3n_2^3 - 3n_3^3 + 2n_4^3 - n_5^3 &=& (2p+1) n_6^3,
\end{eqnarray}
where an extra right-handed fermion multiplet transforming as $(1,2p+1,0;n_6)$ 
has been added to each family of quarks and leptons.

To find solutions to the above 3 equations, consider first $p=0$, then 
Eq.~(16) forces one to choose solution (I), and all other equations are 
satisfied with $n_1=n_2=n_3$ and $n_4=n_5=n_6$, i.e. $U(1)_{B-L}$ has been 
obtained.  Consider now $p \neq 0$, then if solution (I) is again 
chosen, $n_6=0$ is required, which leads to $U(1)_Y$, so there is nothing new.

Now consider $p \neq 0$ and solution (II).  From Eqs.~(16), (17), and (18), 
it is easily shown that
\begin{equation}
{4n_6 \over 3n_1+n_4} = {6 \over p(p+1)(2p+1)} = {3 \over 2p+1} = 
\left( {3 \over 2p+1} \right)^{1\over 3},
\end{equation}
which clearly gives the unique solution of $p=1$, i.e. a triplet.  The fact 
that such a solution even exists (and for an integer value of $p$) for the 
above overconstrained set of conditions is certainly not a ``trivial'' or 
even ``expected'' result.

The $U(1)_X$ charges of the possible Higgs doublets are:
\begin{equation}
n_1-n_3 = n_2-n_1 = n_6-n_4 = {3 \over 4} (n_1-n_4), ~~~ n_4-n_5 = {1 \over 4} 
(9n_1-n_4),
\end{equation}
which means that two distinct Higgs doublets are sufficient for all possible 
Dirac fermion masses in this model.  If $n_4 = -3n_1$ is chosen, then again 
$U(1)_X$ will be proportional to $U(1)_Y$.  However, for $n_4 \neq -3n_1$, 
a new class of models is now possible with $U(1)_X$ as a genuinely new gauge 
symmetry.

\section{Scenario A: Neutrino-Quark Scattering}

Consider $\nu q$ and $\bar \nu q$ deep inelastic scattering.  It has 
recently been reported \cite{nutev} by the NuTeV Collaboration that their 
measurement of the effective $\sin^2 \theta_W$, i.e. $0.2277 \pm 0.0013 \pm 
0.0009$, is about $3\sigma$ away from the standard-model prediction of 
$0.2227 \pm 0.00037$.  In this model, the $X$ gauge boson also contributes 
with
\begin{eqnarray}
J_X^\mu &=& n_1 \bar u \gamma^\mu \left( {1-\gamma_5 \over 2} \right) u + 
n_1 \bar d \gamma^\mu \left( {1-\gamma_5 \over 2} \right) d \nonumber \\ 
&& +~n_2 \bar u \gamma^\mu \left( {1+\gamma_5 \over 2} \right) u + 
n_3 \bar d \gamma^\mu \left( {1+\gamma_5 \over 2} \right) d + n_4 \bar \nu 
\gamma^\mu \left( {1-\gamma_5 \over 2} \right) \nu.
\end{eqnarray}
Assuming very small $X-Z$ mixing ($|\sin \theta| << 1$), the effective 
neutrino-quark interactions are then given by
\begin{equation}
{\cal H}_{int} = {G_F \over \sqrt 2} \bar \nu \gamma^\mu (1-\gamma_5) \nu 
[\epsilon_L^q \bar q \gamma_\mu (1-\gamma_5) q + \epsilon_R^q \bar q 
\gamma_\mu (1+\gamma_5) q],
\end{equation}
where
\begin{eqnarray}
\epsilon_L^u &=& (1-\xi)\left( {1 \over 2} - {2 \over 3} \sin^2 \theta_W 
\right) + n_1 \zeta, \\ 
\epsilon_L^d &=& (1-\xi)\left( -{1 \over 2} + {1 \over 3} \sin^2 \theta_W 
\right) + n_1 \zeta, \\ 
\epsilon_R^u &=& (1-\xi)\left( -{2 \over 3} \sin^2 \theta_W \right) 
+ n_2 \zeta, \\ 
\epsilon_R^d &=& (1-\xi)\left( {1 \over 3} \sin^2 \theta_W \right) 
+ n_3 \zeta,
\end{eqnarray}
with
\begin{eqnarray}
\xi &=& 2 n_4 \sin \theta \left( 1 - {M_Z^2 \over M_X^2} \right) {g_X \over 
g_Z}, \\ 
\zeta &=& - \sin \theta \left( 1 - {M_Z^2 \over M_X^2} \right) {g_X \over g_Z} 
+ 2 n_4 \left( {M_Z^2 \over M_X^2} \right) {g_X^2 \over g_Z^2}.
\end{eqnarray}

The parameter $\xi$ is constrained by data at the $Z$ resonance to be very 
small.  Using the general analysis of $Z-X$ mixing \cite{abj}, we find
\begin{eqnarray}
\xi &=&  {2 s^2 c^2 \over c^2-s^2} \Delta \epsilon_1 + 
 {(c^2-s^2)^2 \over 2 c^2} \Delta \epsilon_2 + {s^2 \over c^2} 
{(-1-2s^2+4s^4) \over c^2-s^2} \Delta \epsilon_3 \nonumber \\ 
&=& 0.624 \Delta \epsilon_1 + 0.198 \Delta \epsilon_2 - 0.644 
\Delta \epsilon_3,
\end{eqnarray}
where $s \equiv \sin \theta_W$ and $c \equiv \cos \theta_W$.  Given that 
$|\Delta \epsilon_i|$ is of order 0.001, $\xi$ is too small to make much 
difference in the above \cite{dfgrs}.  We thus assume $\xi = 0$ ($\sin \theta 
=0$) for our subsequent discussion.

To account for the NuTeV result, i.e.
\begin{eqnarray}
(g_L^{eff})^2 &=& (\epsilon_L^u)^2 + (\epsilon_L^d)^2 = 0.3005 \pm 0.0014, \\ 
(g_R^{eff})^2 &=& (\epsilon_R^u)^2 + (\epsilon_R^d)^2 = 0.0310 \pm 0.0011,
\end{eqnarray}
against the standard-model prediction, i.e.
\begin{equation}
(g_L^{eff})^2_{SM} = 0.3042, ~~~ (g_R^{eff})^2_{SM} = 0.0301,
\end{equation}
consider the following specific model as an illustration:
\begin{equation}
n_1 = 1, ~~ n_2 = {3 \over 4}, ~~ n_3 = {5 \over 4}, ~~ n_4 = {4 \over 3}, ~~ 
n_5 = -{7 \over 12}, ~~ n_6 = {13 \over 12}.
\end{equation}
Then
\begin{eqnarray}
\Delta (g_L^{eff})^2 &=& -{2 \over 3} \sin^2 \theta_W \zeta + 2 \zeta^2, \\ 
\Delta (g_R^{eff})^2 &=& -{1 \over 6} \sin^2 \theta_W \zeta + {17 \over 8} 
\zeta^2.
\end{eqnarray}
To fit the experimental values, we need a negative $\Delta (g_L^{eff})^2$. 
From Eq.~(34) we see that it reaches its maximum value at
\begin{equation}
\zeta = {1 \over 6} \sin^2 \theta_W,
\end{equation}
for which
\begin{eqnarray}
\Delta (g_L^{eff})^2 &=& -{1 \over 18} \sin^4 \theta_W = -0.0028, \\ 
\Delta (g_R^{eff})^2 &=& {1 \over 32} \sin^4 \theta_W = +0.0016,
\end{eqnarray}
in very good agreement with the experimental values of $-0.0037 \pm 0.0014$ 
and $+0.0009 \pm 0.0011$ respectively.

Using Eqs.~(28), (33), and (36), we find that 
\begin{equation}
{g_X^2 \over M_X^2} = {\sin^2 \theta_W \over 16} {g_Z^2 \over M_Z^2}.
\end{equation}
Thus the production of the new gauge boson $X$ may be studied as a function 
of the single parameter $M_X$ in this scenario.  We note first that if the 
$U(1)_X$ assignments of Eq.~(33) apply to electrons as well, then Eq.~(39) 
is in serious conflict with atomic parity violation and $e^+ e^-$ cross 
sections.  We must therefore attribute the NuTeV anomaly as being due to the 
muon (and perhaps also the tau) sector \cite{dfgrs,maroy} only.  In the 
context of $U(1)_X$, this may be accomplished as follows.  We change the 
electron's assignments under $U(1)_X$ to zero so that it does not couple to 
$X$ at all. To preserve the cancellation of anomalies, we add heavy fermions 
at the TeV energy scale, i.e.
\begin{eqnarray}
&& (N,F)_L \sim (1,2,-1/2;n_4), ~~~(N,F)_R \sim (1,2,-1/2;0); \\
&& E_L \sim (1,1,-1;0), ~~~E_R \sim (1,1,-1;n_5).
\end{eqnarray}
These are prevented from coupling to the known leptons by a discrete symmetry 
to forbid terms such as $\bar E_L e_R$, etc.  As a result, the lightest among 
them is stable, in analogy to the lightest supersymmetric 
particle of $R$-parity conserving supersymmetry.

The spin and color averaged matrix element square for the $X$ boson signal at 
the Tevatron and at the LHC is given by
\begin{equation}
\overline {M^2}_{q \bar q \to X \to f \bar f} = {1 \over 3} {g_X^4 \over 
(s-M_X^2)^2 + M_X^2 \Gamma_X^2} \left[ n^2_{q_L} (u^2 n^2_{f_L} + t^2 
n^2_{f_R}) + n^2_{q_R} (u^2 n^2_{f_R} + t^2 n^2_{f_L}) \right],
\end{equation}
where
\begin{equation}
\Gamma_X = {g_X^2 \over 24 \pi} M_X (18 n_1^2 + 9 n_2^2 + 9 n_3^2 + 4 n_4^2 + 
2 n_5^2).
\end{equation}
Substituting the required value of $g_X$ from Eq.~(39) and the $X$ charges 
($n_i$) from Eq.~(33) we see that for $M_X > 1$ TeV, $g_X > 1$ and its width 
becomes comparable to its mass.  Figure 3 shows the total $X$ boson 
production cross sections at the LHC and at the Tevatron as functions of its 
mass.  We see that the predicted signal cross sections are really large if 
the $X$ boson is to account for the NuTeV anomaly.  It may be noted here 
that there is a 95\% confidence-level upper limit of
\begin{equation}
\sigma (X) B(X \to e^+ e^- \& \mu^+ \mu^-) = 40 ~fb
\end{equation}
from the CDF experiment \cite{cdf} at the Tevatron.  The $X$ charges of 
Eq.~(33) correspond to a branching fraction $B(X \to \mu^+ \mu^-) = 4-5$ \%. 
Thus assuming the CDF detection efficiency to be roughly similar for the 
$e^+ e^-$ and $\mu^+ \mu^-$ channels, the above constraint would imply 
$\sigma(X) < 1000 ~fb$ at the Tevatron.  On the other hand we see from 
Figure 3 that $\sigma(X) > 1000 ~fb$ at the Tevatron right up to $M_X = 2$ 
TeV (the finite value at the kinematic boundary is due to the large width).  
Thus consistency with this limit will require $M_X > 2$ TeV.  Figure 2a shows 
a very large signal cross section at the LHC up to $M_X = 3$ TeV, 
corresponding to $g_X \simeq 3$.  It remains large at larger values of $M_X$ 
as well, the cutoff being provided by the perturbation theory limit on $g_X$.

\section{Scenario B:  Atomic Parity Violation}

Instead of trying to accommodate the NuTeV discrepancy, we now consider the 
possibility of having a small effect in atomic parity violation from 
$U(1)_X$.  Using the most recent precise atomic calculation of the 6s--7s 
parity violating E1 transition in cesium \cite{apv}, a slight deviation 
from the standard-model prediction is obtained, i.e.
\begin{equation}
\Delta Q_W = 0.91 \pm 0.29 \pm 0.36.
\end{equation}
In the $U(1)_X$ model, with the definition $r \equiv n_1/n_4$, we have 
\cite{bcrz}
\begin{equation}
\Delta Q_W = -376 C_{1u} - 422 C_{1d} = {3 \over 4} {g_X^2 \over M_X^2} 
{M_Z^2 \over g_Z^2} (1041r + 23)(1 - 9r).
\end{equation}
This shows that there is only a narrow range for $r$ which yields a positive 
value of $\Delta Q_W$.  For illustration, we choose $r=0$, i.e.
\begin{equation}
n_1 = 0, ~~n_2 = -{3 \over 4}, ~~n_3 = {3 \over 4}, ~~n_4 = 1, ~~n_5 = 
{5 \over 4}, ~~n_6 = {1 \over 4}.
\end{equation}
From Eqs.~(45) and (46), we then find
\begin{equation}
0.015 < {g_X^2 \over M_X^2} {M_Z^2 \over g_Z^2} < 0.09.
\end{equation}
Since this range of values will lead to large effects in the muon sector 
if $X$ couples in the same way, we assume in this scenario that $X$ 
couples only to electrons and quarks.  By doing so, we also avoid serious 
constraints from $e^+ e^- \to \mu^+ \mu^-, \tau^+ \tau^-$ measurements 
at LEP2 \cite{aleph,pdg}.

We have estimated the $X$ boson production cross sections at the Tevatron and 
at the LHC using Eq.~(42) for the lower limit of the $X$ coupling from 
Eq.~(48), which is practically identical to that of Eq.~(39).  They are 
shown in Figure 4.  They are somewhat smaller than those of Figure 3 due 
to the different $X$ charges.  On the other hand, we expect a high $B(X \to 
e^+ e^-) \simeq 18$ \% in this case.  Thus the CDF limit of Eq.~(44) would 
again imply $M_X > 2$ TeV.  Nonetheless we expect a very large signal cross 
section at the LHC up to a mass range of several TeV, till the value of 
$g_X$ is again cut off by the perturbation theory limit.

\section{Higgs Sector}

As shown in Sec.~3, $U(1)_X$ requires two distinct Higgs doublets for fermion 
masses, i.e. $\Phi_1 = (\phi_1^+, \phi_1^0)$ with $U(1)_X$ charge 
$(9n_1-n_4)/4$ which couples to charged leptons, and $\Phi_2 = (\phi_2^+, 
\phi_2^0)$ with $U(1)_X$ charge $(3n_1-3n_4)/4$ which couples to $up$ and 
$down$ quarks as well as to $\Sigma$.  [The leptonic Higgs doublet $\eta$ 
of Sec.~2 may also be introduced so that only it couples to $\Sigma$, while 
$\Phi_2$ couples only to quarks.]  To break the $U(1)_X$ gauge symmetry 
spontaneously, we add a singlet $\chi$ with $U(1)_X$ charge $-2n_6$, so 
that the Yukawa term $\chi \Sigma \Sigma$ would allow $\Sigma$ to acquire a 
large Majorana mass at the $U(1)_X$ breaking scale. The Higgs potential of 
this model is then given by
\begin{eqnarray}
V &=& m_1^2 \Phi_1^\dagger \Phi_1 + m_2^2 \Phi_2^\dagger \Phi_2 + m_3^2 
\chi^\dagger \chi + [f \chi^\dagger \Phi_1^\dagger \Phi_2 + h.c.] \nonumber \\ 
&+& {1 \over 2} \lambda_1 (\Phi_1^\dagger \Phi_1)^2 + {1 \over 2} \lambda_2 
(\Phi_2^\dagger \Phi_2)^2 + {1 \over 2} \lambda_3 (\chi^\dagger \chi)^2 
+ \lambda_4 (\Phi_1^\dagger \Phi_1)(\Phi_2^\dagger \Phi_2) \nonumber \\ 
&+& \lambda_5 (\Phi_1^\dagger \Phi_2)(\Phi_2^\dagger \Phi_1) + \lambda_6 
(\chi^\dagger \chi)(\Phi_1^\dagger \Phi_1) + \lambda_7 (\chi^\dagger \chi) 
(\Phi_2^\dagger \Phi_2).
\end{eqnarray}
Note that the $U(1)_X$ charges allow the trilinear term $\chi^\dagger 
\Phi_1^\dagger \Phi_2$, without which $V$ would have 3 global U(1) symmetries, 
but only 2 U(1) gauge symmetries, resulting in an unwanted Goldstone boson.
We have not included $\eta$ because $m_\eta^2$ is large and positive as 
discussed in Sec.~2.  After the heavy $\chi$ [with $\langle \chi \rangle \sim 
1$ TeV] has been integrated out, the reduced two-doublet Higgs potential is 
of the usual form.  The difference from other proposals is in their Yukawa 
couplings, i.e. $\Phi_1$ couples to charged leptons whereas $\Phi_2$ couples 
to both $up$ and $down$ quarks.

Let us briefly discuss the distinctive phenomenological features of this 
two-Higgs-doublet model.  While the vacuum expectation value $\langle \phi_2^0 
\rangle$ is required to be $\sim 100$ GeV because of $m_t$, $\langle \phi_1^0 
\rangle$ can be anywhere between $\sim 100$ GeV and 1--2 GeV.  In terms of the 
ratio $\tan \beta = \langle \phi_2^0 \rangle/\langle \phi_1^0 \rangle$, they 
corespond to the limits $\tan \beta \simeq 1$ and $\tan \beta >> 1$.  In the 
former case, the phenomenological implications are similar to those of the 
standard two-Higgs-doublet scenario where $\Phi_2$ couples to the $up$ quarks 
and $\Phi_2$ couples to the charged leptons as well as to the $down$ quarks.  
In the latter case however, there are distinctive differences between our 
proposal and the standard scenario, because Higgs couplings proportional to 
$m_b$ are now mutliplied by $\cot \beta$ instead of $\tan \beta$.  Let us 
consider in particular the charged and the pseudoscalar neutral Higgs bosons, 
$H^\pm$ and $A^0$, which correspond to the linear combination $\Phi_1 \sin 
\beta - \Phi_2 \cos \beta$.  Their distinctive phenomenological features are 
summarized below.

i) The $H^- t$ loop contribution to the $b \to s \gamma$ decay amplitude 
is suppressed.  It has the factor $\cot^2 \beta$ instead of $\tan \beta 
\cot \beta = 1$ in the standard scenario.  This means that the charged Higgs 
boson in this model can be relatively light.

ii) The $H^- \to \tau^- \bar \nu$ decay dominates over $H^- \to \bar t b$ 
for any charged Higgs mass.

iii) The $H^\pm$ production via $t \bar b$ fusion is no longer the main 
production mechanism at hadron colliders.  It will instead be pair production 
via the Drell-Yan mechanism discussed in Section 2.  The plots of Figure 1 
apply equally to $H^\pm$ pair production in this case.  Thus we expect a 
visible signal at the LHC up to a Higgs mass of about 1 TeV.

iv) The $A^0 \to \tau^+ \tau^-$ decay dominates over $A^0 \to b \bar b$ and 
$A^0 \to t \bar t$.

v) The $A^0$ production is again no longer dominated by $b \bar b$ or $t 
\bar t$ fusion at hadron colliders, but rather by associated production of 
$A^0 H^0 (H^\pm)$ through $Z (W^\pm)$ exchange.

There are analogous distinctions for the two physical neutral scalars $h^0$ 
and $H^0$.  However, they depend on the additional mixing angle $\alpha$ 
which is not necessarily close to $\beta$.

\section{Conclusion}

In conclusion, we have elaborated on a recently proposed model \cite{ma02} 
of neutrino mass, where a heavy triplet of leptons $(\Sigma^+, \Sigma^0, 
\Sigma^-)_R$ per family is added as the anchor of the seesaw mechanism 
in place of the canonical singlet $N_R$.  Using a second ``leptonic'' Higgs 
doublet $\eta$ \cite{ma01}, both $M_\Sigma$ and $m_\eta$ can be at the TeV 
scale and be produced at the LHC.  Experimental determination of their masses 
and decays to charged leptons will then map out the neutrino mass matrix. 

The existence of the triplet $\Sigma$ allows a new U(1) gauge symmetry 
\cite{ma02} to be defined, with the associated gauge boson $X$ at or 
below the TeV scale.  The $U(1)_X$ charges of the usual quarks and 
leptons are fixed up to one free parameter.  If $g_X^2/M_X^2$ is not too 
small compared to $g_Z^2/M_Z^2$, there will be corrections to the standard 
model at low energies coming from this new gauge symmetry.  We consider 
two scenarios.  (A) The recent NuTeV anomaly \cite{nutev} is explained 
by $U(1)_X$.  (B) The possible slight discrepancy \cite{apv} in atomic parity 
violation \cite{bewi} is explained by $U(1)_X$.  We find that (A) and (B) 
are mutually exclusive, i.e. we can accommodate one but not the other. 
The resulting constraint from either (A) or (B) on $g_X^2/M_X^2$ is such 
that the production of $X$ has to be very large at hadron colliders. 
Present limits at the Tevatron imply that $M_X$ is larger than about 2 TeV. 
On the other hand, it is quite possible that these anomalies have other 
explanations, in which case there is no hard constraint on $M_X$ or $g_X$.  
Nevertheless, for $g_X$ of order the electroweak coupling, the production 
of $M_X$ up to a few TeV can be reached at the LHC.

The $U(1)_X$ gauge symmetry requires two Higgs doublets of different 
$U(1)_X$ charges, such that $\Phi_1$ couples to charged leptons and 
$\Phi_2$ couples to both $up$ and $down$ quarks.  This differs from the 
standard scenario and allows in particular the charged Higgs boson to be 
light in the case of large $\tan \beta$, resulting in a number of distinct 
phenomenological predictions.

This work was supported in part by the U.~S.~Department of Energy under 
Grant No.~DE-FG03-94ER40837.

\newpage
\bibliographystyle{unsrt}

\newpage

\begin{figure}
\begin{center}
\vspace*{4.5in}
\includegraphics{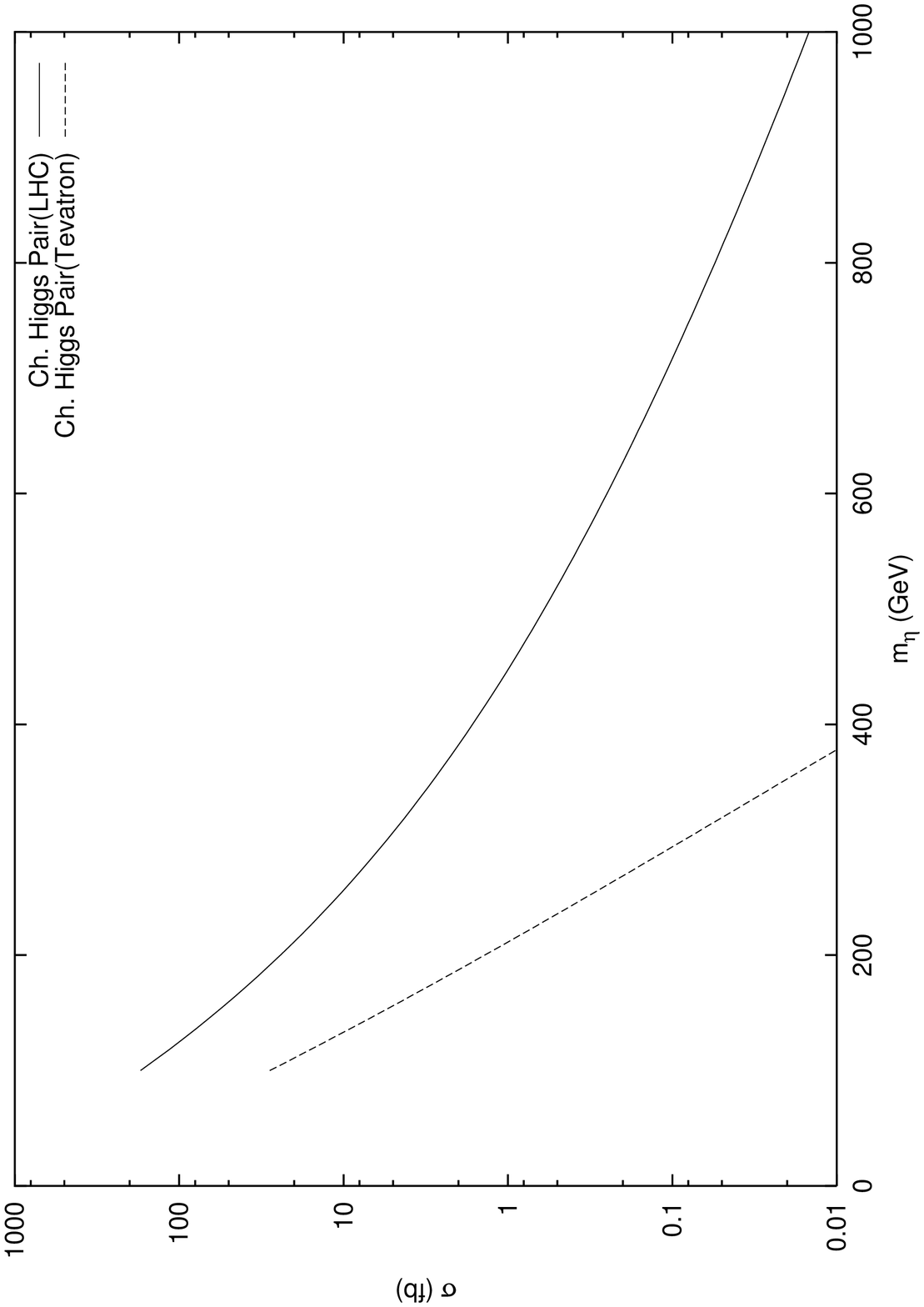}
\end{center}
\caption{Cross sections for $\eta^\pm$ pair production at LHC and Tevatron.}
\end{figure}

\newpage

\begin{figure}
\begin{center}
\vspace*{4.5in}
\includegraphics{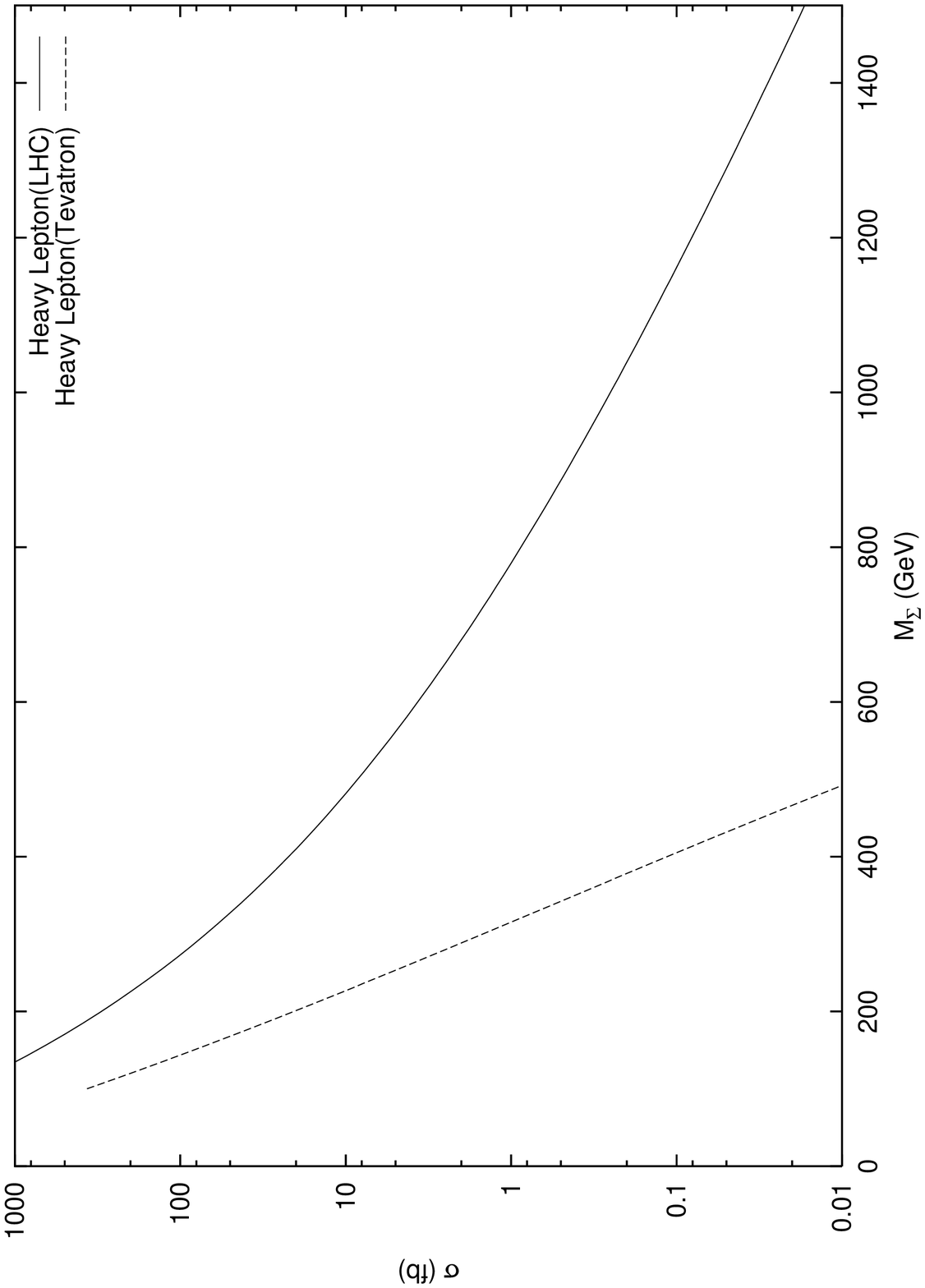}
\end{center}
\caption{Cross sections for $\Sigma^\pm$ pair production at LHC and Tevatron.}
\end{figure}

\newpage

\begin{figure}
\begin{center}
\vspace*{4.5in}
\includegraphics{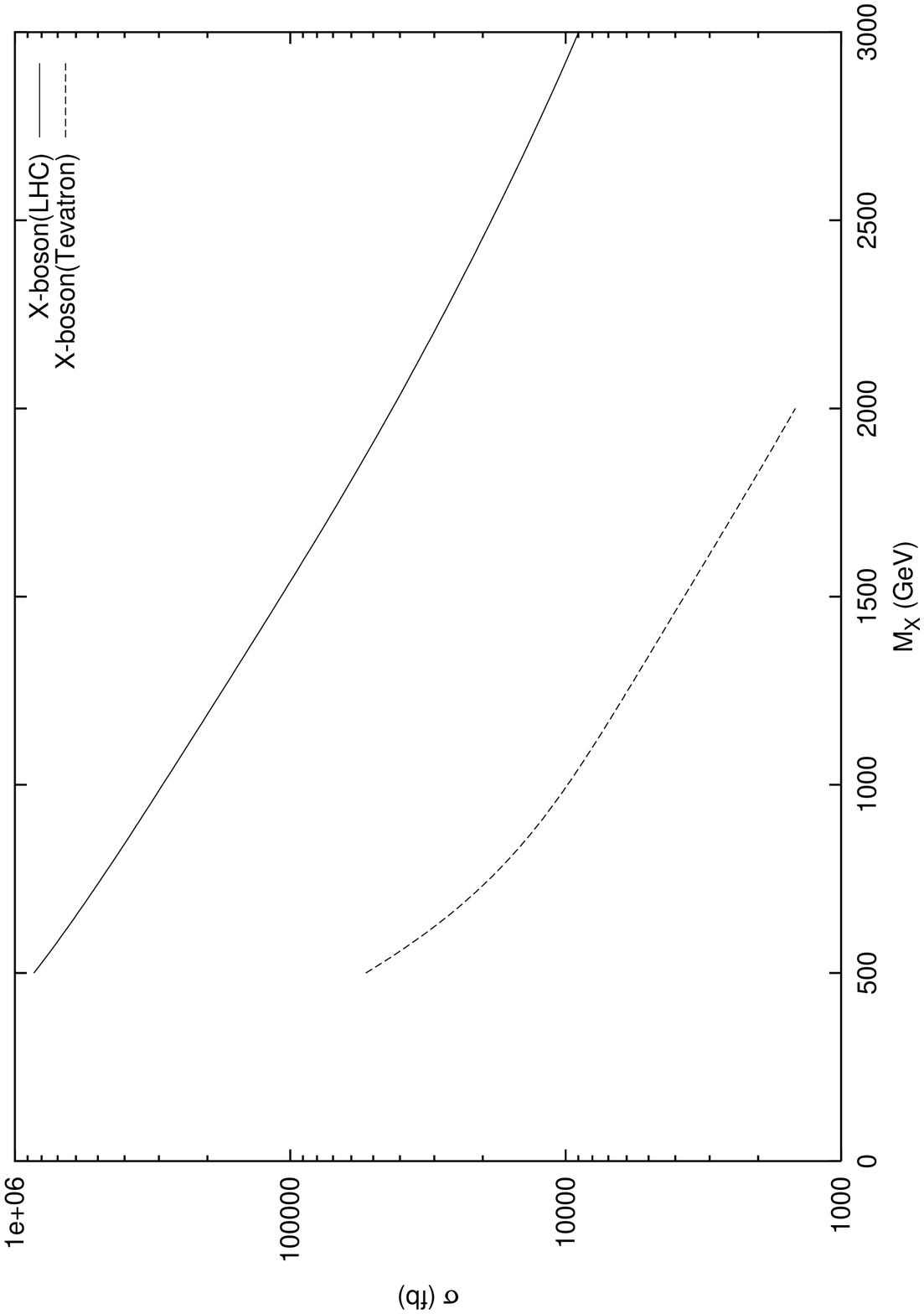}
\end{center}
\caption{Cross sections for $X$ production at LHC and Tevatron in Scenario A.}
\end{figure}

\newpage

\begin{figure}
\begin{center}
\vspace*{4.5in}
\includegraphics{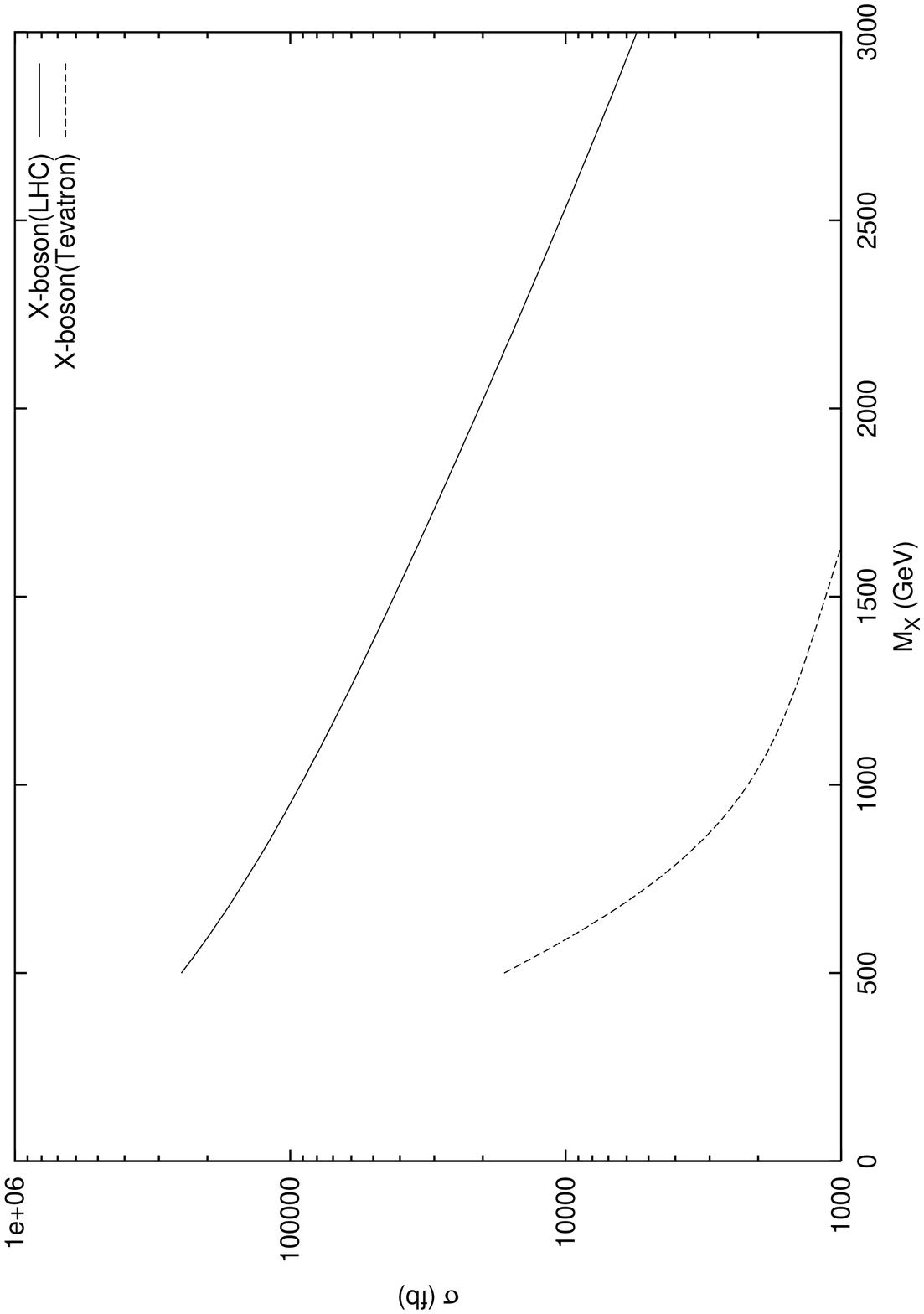}
\end{center}
\caption{Cross sections for $X$ production at LHC and Tevatron in Scenario B.}
\end{figure}

\end{document}